\documentclass{aip-cp}
\pdfoutput=1
\usepackage[numbers]{natbib}
\usepackage{rotating}
\usepackage{graphicx}
\usepackage{lineno}

\begin{document}

\newcommand{\nonubb}  {$0 \nu \beta \beta$}
\newcommand{\twonubb}{$2 \nu \beta \beta$}
\newcommand{\bb}{$\beta\beta$}
\newcommand{\onecpRty} {\hbox{1~count/ROI/t-y}}
\newcommand{\MJ}{{\sc{Majo\-ra\-na}}}
\newcommand{\Dem}{{\sc{Demonstrator}}}
\newcommand{\MJD}{{\sc{Majorana Demonstrator}}}
\def\ppc{P-PC}  
\newcommand{\gam}{$\gamma$}
\def\nuc#1#2{${}^{#1}$#2}
\def\cpRty{c/(ROI t yr)}

\title{Contamination Control and Assay Results for the {\sc Majorana Demonstrator} Ultra Clean Components}

\author[sdsmt]{C.D.~Christofferson\corref{cor1}}
\author[lbnl]{N.~Abgrall}
\author[uw]{S.I.~Alvis}
\author[pnnl]{I.J.~Arnquist}
\author[usc,ornl]{F.T.~Avignone~III}
\author[ITEP]{A.S.~Barabash}
\author[usd]{C.J.~Barton}
\author[ornl]{F.E.~Bertrand}
\author[mpi]{T.~Bode}
\author[lbnl]{A.W.~Bradley}
\author[JINR]{V.~Brudanin}
\author[duke,tunl]{M.~Busch}
\author[uw]{M.~Buuck}
\author[unc,tunl]{T.S.~Caldwell}
\author[lbnl]{Y-D.~Chan}
\author[lanl]{P.-H.~Chu}
\author[uw,clara]{C. Cuesta}
\author[uw]{J.A.~Detwiler}
\author[sdsmt]{C. Dunagan}
\author[ut,ornl]{Yu.~Efremenko}
\author[ou]{H.~Ejiri}
\author[lanl]{S.R.~Elliott}
\author[unc,tunl]{T.~Gilliss}
\author[princeton]{G.K.~Giovanetti}
\author[ncsu,tunl,ornl]{M.P.~Green}
\author[uw]{J. Gruszko}
\author[uw]{I.S.~Guinn}
\author[usc]{V.E.~Guiseppe}
\author[unc,tunl]{C.R.~Haufe}
\author[lbnl]{L.~Hehn}
\author[unc,tunl]{R.~Henning}
\author[pnnl]{E.W.~Hoppe}
\author[unc,tunl]{M.A.~Howe}
\author[blhill]{K.J.~Keeter}
\author[ttu]{M.F.~Kidd}
\author[ITEP]{S.I.~Konovalov}
\author[pnnl]{R.T.~Kouzes}
\author[ut]{A.M.~Lopez}
\author[queens]{R.D.~Martin}
\author[lanl]{R. Massarczyk}
\author[unc,tunl]{S.J.~Meijer}
\author[mpi,tum]{S.~Mertens}
\author[lbnl]{J.~Myslik}
\author[unc,tunl]{C. O'Shaughnessy}
\author[unc,tunl]{G.~Othman}
\author[lbnl]{A.W.P.~Poon}
\author[ornl]{D.C.~Radford}
\author[unc,tunl]{J.~Rager}
\author[unc,tunl]{A.L.~Reine}
\author[lanl]{K.~Rielage}
\author[uw]{R.G.H.~Robertson}
\author[uw]{N.W.~Rouf}
\author[unc,tunl]{B.~Shanks}
\author[JINR]{M.~Shirchenko}
\author[sdsmt]{A.M.~Suriano}
\author[usc]{D.~Tedeschi}
\author[unc,tunl]{J.E.~Trimble}
\author[ornl]{R.L.~Varner}
\author[JINR]{S. Vasilyev}
\author[lbnl]{K.~Vetter}
\author[unc,tunl]{K.~Vorren}
\author[lanl]{B.R.~White}
\author[unc,tunl,ornl]{J.F.~Wilkerson}
\author[usc]{C. Wiseman}
\author[usd]{W.~Xu}
\author[JINR]{E.~Yakushev}
\author[ornl]{C.-H.~Yu}
\author[ITEP]{V.~Yumatov}
\author[JINR]{I.~Zhitnikov}
\author[lanl]{B.X.~Zhu}

\affil[sdsmt]{South Dakota School of Mines and Technology, Rapid City, SD, USA}
\affil[lbnl]{Nuclear Science Division, Lawrence Berkeley National Laboratory, Berkeley, CA, USA}
\affil[uw]{Center for Experimental Nuclear Physics and Astrophysics, and Department of Physics, University of Washington, Seattle, WA, USA}
\affil[pnnl]{Pacific Northwest National Laboratory, Richland, WA, USA}
\affil[usc]{Department of Physics and Astronomy, University of South Carolina, Columbia, SC, USA}
\affil[ornl]{Oak Ridge National Laboratory, Oak Ridge, TN, USA}
\affil[ITEP]{National Research Center ``Kurchatov Institute'' Institute for Theoretical and Experimental Physics, Moscow, Russia}
\affil[usd]{Department of Physics, University of South Dakota, Vermillion, SD, USA} 
\affil[mpi]{Max-Planck-Institut f\"{u}r Physik, M\"{u}nchen, Germany}
\affil[JINR]{Joint Institute for Nuclear Research, Dubna, Russia}
\affil[duke]{Department of Physics, Duke University, Durham, NC, USA}
\affil[tunl]{Triangle Universities Nuclear Laboratory, Durham, NC, USA}
\affil[unc]{Department of Physics and Astronomy, University of North Carolina, Chapel Hill, NC, USA}
\affil[lanl]{Los Alamos National Laboratory, Los Alamos, NM, USA}
\affil[ut]{Department of Physics and Astronomy, University of Tennessee, Knoxville, TN, USA}
\affil[ou]{Research Center for Nuclear Physics, Osaka University, Ibaraki, Osaka, Japan}
\affil[princeton]{Department of Physics, Princeton University, Princeton, NJ, USA}
\affil[ncsu]{Department of Physics, North Carolina State University, Raleigh, NC, USA}	
\affil[blhill]{Department of Physics, Black Hills State University, Spearfish, SD, USA}
\affil[ttu]{Tennessee Tech University, Cookeville, TN, USA}
\affil[queens]{Department of Physics, Engineering Physics and Astronomy, Queen's University, Kingston, ON, Canada} 
\affil[tum]{Physik Department, Technische Universit\"{a}t, M\"{u}nchen, Germany}
\affil[clara]{Present Address: Centro de Investigaciones Energ\'{e}ticas, Medioambientales y Tecnol\'{o}gicas, CIEMAT, 28040, Madrid, Spain}
\corresp[cor1]{Corresponding author: CabotAnn.Christofferson@sdsmt.edu}

\maketitle


\begin{abstract}
The \MJD\ is a neutrinoless double beta decay experiment utilizing enriched Ge-76 detectors in 2 separate modules inside of a common solid shield at the Sanford Underground Research Facility. The \Dem\ has utilized world leading assay sensitivities to develop clean materials and processes for producing ultra-pure copper and plastic components. This experiment is now operating, and initial data provide new insights into the success of cleaning and processing. Post production copper assays after the completion of Module 1 showed an increase in U and Th contamination in finished parts compared to starting bulk material. A revised cleaning method and additional round of surface contamination studies prior to Module 2 construction have provided evidence that more rigorous process control can reduce surface contamination. This article describes the assay results and discuss further studies to take advantage of assay capabilities for the purpose of maintaining ultra clean fabrication and process design.
\end{abstract}

\section{INTRODUCTION}
The \MJ\ Collaboration \cite{abg14} is conducting R\&D towards a ton-scale, $^{76}$Ge-based neutrinoless double-beta decay (\nonubb) experiment . The observation of this rare decay would imply the existence of a lepton number violating process and the Majorana nature of neutrinos. The goal for a next-generation \nonubb\ experiment is to reach decay half lives of over 10$^{27}$ yr in order to reach the mass scale in the inverted neutrino mass ordering region. For an experiment to have discovery potential at the level of this goal requires excellent detector energy resolution and background levels in the neutrinoless double-beta decay region of interest (ROI) on the order of 0.1 \cpRty. 

\subsection{\MJD\ Overview}
The \MJ\ Collaboration is operating a \Dem\ array \cite{abg14,ell16} at the Sanford Underground Research Facility, in Lead, SD, USA. The \Dem\ contains two separate detector modules containing a total of 44.1 kg of Ge detectors of which 29.7 kg of the detector mass is enriched to 88\% in $^{76}$Ge \cite{abg17b}. The two modules are operated inside a common low background passive shield with active muon veto (see Ref. \cite{abg14} for a more detailed description of the \MJD). The main goal of the \Dem\ are to show that a background level of 3  \cpRty\ after analysis cuts at the \nonubb\ peak energy can be reached
Reaching the background goal of the \MJD\ will establish that the material purity of the detector and innermost parts are appropriate for a next generation experiment where additional material processing control, background rejection, and improved shielding techniques can be employed.

\subsection{Contamination Control}
The \MJ\ Collaboration conducted a thorough material assay \cite{abg16a} and Monte Carlo simulation program to ensure that all components used in the construction of the \MJD\ array have sufficient radio-purity levels to meet the experiment's background goal. A listing of all simulated background contributions can be found in Table \ref{tab:BG}, which is categorized by the source of the background, for a total background projection of $\leq3.5$ \cpRty.
The main challenge of the assay program is to both identify materials with acceptable natural radioactivity levels of U and Th decay chains and control the handling of the materials to maintain its purity. Through the assay program and additional assay campaigns described here, the Collaboration reached its targeted goal, but we recognize that further improvements can be made when moving towards the next generation experiments. Looking towards a large-scale experiment, lowering the backgrounds intrinsic to detector materials will strengthen the discovery potential for \nonubb. While improving background rejection techniques is important, equally so is contamination control to approach a background-free experiment.  

\begin{table}[h]
\caption{The summary of all the backgrounds contributing to the \Dem. The background contributions are organized by those due to natural radioactivity, cosmogenic activation, external \& environmental, muon-induced, and neutrinos. The background values assume that the radioactive chains are in equilibrium. Adapted from Ref. \cite{abg16a}}
\label{tab:BG}
\tabcolsep7pt\begin{tabular}{lr|lr}
\hline
{\bf Background Contribution}						& {\bf Rate}  &
{\bf Background Contribution}					& {\bf Rate} \\
										& \cpRty  &
										& \cpRty  \\
\hline
Electroformed Cu							& 	0.23	&
OFHC Cu Shielding							&	0.29	\\
Pb Shielding								&      0.63	&
Cables and Internal Connectors				&$<$0.38	\\
Front Ends								&	0.6\phantom{0}	&
U/Th within the Ge							&$<$0.07	\\
Plastics + Other							&	0.39	&&\\
\hline
\nuc{68}{Ge}, \nuc{60}{Co} within the \nuc{enr}{Ge}	&	0.07	&
\nuc{60}{Co} within the Cu						&	0.09	\\
\hline
External $\gamma$ rays, (alpha,n) Reactions		&	0.1\phantom{0}	&
Rn and Surface $\alpha$ Emission				&	0.05	\\
\hline
Ge, Cu, Pb (n,n'gamma) Reactions				&	0.21 &
Ge(n,n')	 Reactions						&	0.17	\\
Ge(n,$\gamma$)							&	0.13	&
Direct $\mu$ Passage						&	0.03 \\
\hline
$\nu$ Induced Background					&$<$0.01	&&\\
\hline 
{\bf Total}									&&	& {\bf $<$3.5\phantom{0}}\\
\hline
\end{tabular}
\end{table}

Fabrication of any component requires strict cleaning and contamination control during production and assembly so as not to recontaminate a pure material.
Copper is the key structural component for the \Dem\ due to its excellent structural and thermal properties, and because its contains no radioactive isotopes. serving as the inner shielding, cryostats, and all detector mounting hardware.  Given the intrinsic high radio purity of copper, the main goal through production (electroforming), machining, and handling is to have the material remain clean.  The copper is electroformed in a clean environment allowing for the reduction of radioactive contaminants from the U and Th radioactive decay chains while conducting the technique underground prevents formation of $^{60}$Co by cosmogenic activation of Cu. Electroforming of the \Dem's copper was completed in two laboratories, separate from the experiment's detector laboratory; at the Ross Campus of the 4850-ft level of SURF and at a shallow underground site at Pacific Northwest National Laboratory (PNNL) in Richland, WA.  The SURF electroforming laboratory ran from July 2011 to Apr. 2016 with ten baths that could hold 316 stainless steel mandrels of various diameters up to 33 cm.  Six additional baths were run at PNNL.  \MJ\ produced  over 40 different electroforms of varying sizes, the largest being $\sim$90 kg at a thickness of 14 mm.  Total electroformed copper produced was over 2500 kg at the average growth rate of 0.033 mm/day.

Once the copper reached the desired thickness, mandrels were transported to the clean machine shop located at the Davis campus of the 4850-ft level adjacent to the detector room.  There the full machine shop was restricted to clean copper, plastics, and approved stainless steel only.  Water based lubricants were the only cutting fluids used if necessary along with deionized water.  All parts generated were uniquely tracked through machining, cleaning and assembly by a custom-built database \cite{orr15} tracking any piece back to the original mandrel and bath chemistry for quality control.
Located inside the \Dem\ detector room at the Davis campus is a dedicated clean wet laboratory operating at better  standards than the outer laboratory space.  This wet laboratory is where all chemical etching, leaching and general cleaning of all parts for the \Dem\ was completed in specially prepared tanks and containers.  Details of all handling and cleaning were updated into the parts tracking database. Once cleaning was completed all parts were stored in nitrogen purge environment to alleviate latent contamination and exposure to radon plate out or any other airborne contaminants.

\section{ASSAY CAMPAIGNS AND RESULTS}
\MJ\ had established that the electroformed copper stock met the radio-purity specifications initially set in the background budget \cite{abg16a}.  Prior to achieving sufficient assay sensitivity, we assayed the copper using ICP-MS at PNNL by sampling the electrolyte in the bath and working backwards to extrapolate the remaining contamination.  This showed good results of the bulk copper  but would not be sufficient for later manufactured parts.  By necessity PNNL developed the world's most sensitive technique for U and Th in copper and plastics validating that the original goal of $<0.3 \mu$Bq/kg for U and Th had been reached.  In fact the bulk copper electroformed for the \Dem\ showed levels $< 0.017 \pm 0.03$ pg $^{238}$U/gCu and $< 0.011 \pm 0.05$ pg $^{232}$Th/gCu, which were the detection limits of the newly developed analytical method.  We understood we had to maintain a series of assay campaigns to verify the finished machined parts remained pure.  The various handling conditions of each type of part introduced unique pathways of surface contamination additions of U/Th that had not been in the bulk electroformed copper pre-machining.  Should contamination occur, we would be most sensitive to the large mass, large surface area components that surround the entire array. The assay campaigns were designed sequentially after each phase to systematically evaluate the potential pathways of surface recontamination during each step of material handling. The remainder of this article details the findings of each assay campaign.

\subsection{Electroformed copper assay campaign 1}
The first assay campaign used bulk copper pieces cut from mandrels after initial machining within the machine shop.  Compared to smaller parts or plates this is a minimal handling scenario and the closest to bulk stock that could be tested.  Copper was removed in test strip samples and sent to PNNL for cleaning, rinsing, etching, and then full digestion.  The etchant and full digestion of the pieces were assayed.  The results (Fig. \ref{fig:assay}) show the stock electroformed copper was at the level ($<0.3 \mu$Bq/kg for U and Th) needed for use in the \Dem.

\subsection{Electroformed copper assay campaign 2}
The second assay campaign used samples with realistic handling scenarios such as those for an original finished part.  These samples had additional machining and contacted a greater variety of tools in the machine shop. The standard etching procedure from PNNL \cite{hop07} for use at SURF was utilized.  After storing in dry nitrogen, parts were sent to PNNL for assay where no additional processing occurred.  Parts were fully digested and the etchant assayed.  This sample group saw elevated surface contamination which would have varying consequences for parts of the \Dem\ depending on which component had contamination.  With the results normalized by mass, not all parts had the same surface area to mass ratio.  The results in Fig. \ref{fig:assay} are given for hollow hex rods (HHR), which are detector string parts with high surface area to mass ratios, and parts cut with the wire electrical discharge machining (EDM) process. . Given the small mass of these parts, the increased radioactive contamination may not alone cause an increase in background. However, if the same mass concentration of contamination also prevailed for the inner copper shield, which had not been processed, the increase in projected backgrounds could be significant. Given the timing of this result, changes to the processing could be implemented before the inner Cu shield was installed in the \MJD.

\subsection{Electroformed copper assay campaign 3}
The third campaign was a two-part study to confirm the previous observations that the contamination had been introduced as a surface versus a bulk effect.  Similar starting test samples were prepared with a variety of machine shop tooling.  The first part of the study was taking multiple standard etches at SURF, collecting the etchant, and sending to PNNL for assay.  The second part was taking the parts previously etched at SURF and sending to PNNL for additional etching then assaying those samples.  These assay sets now only looked at the etchant left from the surfaces rather than a full digestion of the parts.  
The higher assay levels suggested processing has introduced all of the contaminants. 
Extrapolating among the results of multiple etches gave the prediction for the necessary depth of copper removal to eliminate the surface contamination added from machining.

Based on the understanding that the contamination may be introduced in fabrication and handling of parts, the cleaning procedure was reviewed for improvements.  All the etching bins for large pieces were again acid etched along with changing of smaller containers to a different unique beaker that had only been exposed to electroformed copper parts and high purity reagents.  A new goal was set to find a sufficient depth of Cu to prevent surface contamination that did not also compromise material dimensional tolerances. 
To confirm this process a method for calibration of etching solutions was devised confirming the etching efficiency along with control of the amount of copper removed.  Further detailed handling requirements at each step were used to avoid cross contamination and additional quality assurance.  The timing of this change in cleaning procedure meant that the inner copper shield and all of Module 2 parts would receive the new updated handling while all of Module 1 had been assembled under the standard method. 

\subsection{Electroformed Copper Assay Campaign 4}
\begin{figure}[t]
\centering
  \includegraphics[width=0.65\textwidth]{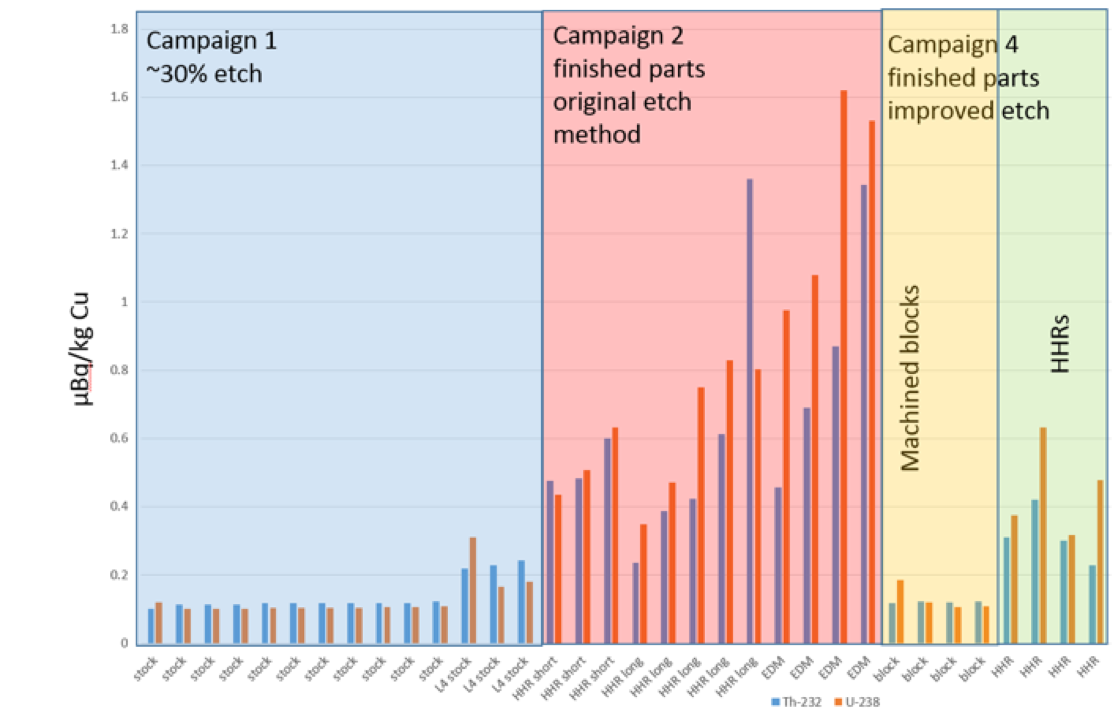} 
  \caption{The ICP-MS results from the four assay campaigns. Each assayed parts is identified wither as being `stock' taken directly from the mandrel, `HHR' a Hollow Hex Rod indicative of a high surface area to mass ratio part, `EDM' for representative parts cut with the Wire Electrical Discharge Machining process, or `block' representative of low surface area to mass ratio parts. The activity of both U and Th is given by color as determined by ICP-MS assay.}
  \label{fig:assay}
\end{figure}

This assay round was anticipated to be validation of the previous three assay campaigns to test if the improved processing and handling method developed had the effect of returning the processed copper contamination to that of the stock values.  The sample group was made up of machined blocks which would be representative of shield plates with low surface area to mass ratio.  The second would be hollow hex rods which are the highest area to mass ratio component of the detector string assemblies.  These two sample groups were etched using the new deeper calibrated method along with altered handling.  The parts were then sent to PNNL for a deeper etch and full digestion assay.  The results (Fig. \ref{fig:assay})  validated that the contamination we found in assay campaign two was likely added during the processing of electroformed copper and that the new cleaning procedure returned the copper to the acceptable purity levels found in the stock material.

\section{OUTLOOK}
The effect on the \MJD\ backgrounds were reviewed in data sets 3 and 4 (DS3 \& 4) which were the latest public release at the time of this presentation.  The exposure was at 1.39 kg y and after cuts equated to 1 count in the 400-keV window centered at the 2039 keV \nonubb peak.  The projected background rate is $5.1^{+8.9}_{-3.2}$ \cpRty\ for a 2.9 (M1) \& 2.6 (M2) keV ROI (68\% CL) with a background index $1.8 \times 10^{-3}$ c/(keV kg yr).  Analysis cuts are still being optimized but are consistent with our expectation of the 3.5 \cpRty\ based on assays \cite{abg16a} and simulations.  The overall handling and processing of the electroformed Cu appears to be acceptable.  Comparing an energy region that would be dominated by U/Th (1950 -- 2650 keV) the collaboration is looking at possible M1 vs.  M2 differences but currently the statistics are too low to make a conclusive statement.  

Building on the success of many low background programs and experiments, the \MJD\ has achieved successful production of ultra-pure underground electroformed copper and an effective initial procedure to handle and process materials to control the surface U/Th contamination of Cu parts.   The parallel assay campaigns systematically evaluated the handling and processing procedure for the range of Cu components and showed a more rigorous etching procedure was successful as confirmed by ICP-MS assay data. With the improved process procedure, the components approach U/Th radio-purity levels consistent with the bulk stock material.  The \Dem's initial backgrounds are among the lowest backgrounds in the region of interest achieved to date (approaching GERDA's recent best value \cite{ago17}).  With ongoing analysis of the current data (10 kg-years), we will be able to make a quantitative statement on the backgrounds in Modules 1 and 2. 

Our assay campaigns continue as we perform additional R\&D on material purity for the next generation \nonubb\ experiments to further systematically explore the source of material contamination. As the \MJD\ continues to take data, we will gain more information on the source of the backgrounds present and couple the results with what is learned from all of the assay campaigns. A newly formed Large Enriched Germanium Experiment for Neutrinoless Double Beta Decay (LEGEND) collaboration builds on the strengths of the GERDA and \MJD\ designs for a ton-scale $^{76}$Ge experiment. LEGEND is estimated to reach a background on the order of 0.1 \cpRty\ so maintaining and controlling material purity, through the practices established here, will be an important component of that goal.

\section{ACKNOWLEDGMENTS}
This material is based upon work supported by the U.S. Department of Energy, Office of Science, Office of Nuclear Physics, the Particle Astrophysics and Nuclear Physics Programs of the National Science Foundation, and the Sanford Underground Research Facility. 

\bibliographystyle{aipnum-cp}%
\bibliography{/Users/guiseppe/work/latex/bib/mymj}%

\end{document}